\documentclass[prl,aps,preprint,amsmath,amssymb]{revtex4}
\usepackage{graphicx}
\usepackage{epsfig}
\include{graphics}

\begin{document}

\title{Scintillation index for two Gaussian laser beams with different 
wavelengths in weak atmospheric turbulence}

\author{Avner Peleg$^{1}$ and Jerome V. Moloney$^{1,2}$}

\affiliation{$^{1}$Arizona Center for  Mathematical Sciences, University of
Arizona, Tucson, Arizona 85721, USA}

\affiliation{$^{2}$College of Optical Sciences, University of Arizona, 
Tucson, Arizona 85721, USA}


\email{avner@acms.arizona.edu}

\begin{abstract}
We study propagation of two lowest order Gaussian laser beams 
with different wavelengths in weak atmospheric turbulence. 
Using the Rytov approximation and assuming 
a slow detector we calculate the longitudinal and radial components of the
scintillation index for a typical free space laser communication setup. 
We find the optimal configuration of the two laser 
beams with respect to the longitudinal scintillation index. We show that the
value of the longitudinal scintillation for the optimal 
two-beam configuration is smaller by more than 50$\%$ 
compared with the value for a single lowest order Gaussian beam 
with the same total power. Furthermore, the radial scintillation for
the optimal two-beam system is smaller by 35$\%$-40$\%$ 
compared with the radial scintillation in the single beam case. 
Further insight into the reduction of intensity fluctuations is 
gained by analyzing the self- and cross-intensity contributions 
to the scintillation index.      
\end{abstract}


\maketitle 

\section{Introduction}
Propagation of light through atmospheric turbulence 
is the subject of a rich and very active field of research owing to 
the many applications in free space 
laser communications, remote sensing, imaging systems and targeting 
\cite{Ishimaru78,Fante85,Andrews98}. 
In these applications it is usually desirable
to find ways to reduce the turbulence effects on the propagating optical
beam. In recent years there has been a renewed interest in using 
partially coherent sources of light as a method for reducing the 
turbulence effects and improving the system performance. 
Many works concentrated on the case where the optical source 
is spatially partially coherent.
First, the properties of the mutual coherence function of 
spatially partially coherent beams propagating in atmospheric 
turbulence were analyzed \cite{Belenkii77,Wang79,Belenkii80}. 
Later on it was shown that intensity fluctuations of spatially 
partially coherent Gaussian beams can decrease as the coherence of 
the sources decreases \cite{Banakh83a,Banakh83b}. Additional 
computational evidence for the smaller sensitivity of spatially 
partially coherent beams to turbulence was obtained by considering
the distance dependence of the coherence radius of the beam 
\cite{Wu90,Wu91}. More recently is was shown 
theoretically \cite{Gbur2002} and experimentally \cite{Dogariu2003} 
that the effective propagation distance measuring beam spreading in 
a turbulent medium relative to free-space spreading is larger for 
spatially partially coherent beams than for fully coherent ones. 
Other works demonstrated the improved performance obtained by 
using spatially partially coherent beams in terms of the normalized 
intensity distribution \cite{Wolf2003a}, 
the mean squared beam width \cite{Wolf2003b,Ji2005}, 
power in the bucket and Strehl ratio \cite{Ji2005}, 
the scintillation index \cite {Andrews2004}, 
the bit error rate (BER)\cite{Ricklin2002,Ricklin2003,Andrews2004} 
and the average signal to noise ratio (SNR)\cite{Andrews2004}.     

Generation of spatially partially coherent beams in various setups, 
e.g., by focusing a laser beam on a rotating random phase screen, is
quite straightforward. However, it is not clear to what extent such
setups would fit in actual applications where reliability and 
compactness of the optical source play an important role. Furthermore,
even though a theoretical method for optimizing a spatially partially
coherent source was outlined in Ref. \cite{Schulz2005}, the problem of
implementing the method for realistic systems remains far from
being resolved. Consequently, one has to look for other methods for
generating the partial coherence, which are potentially 
more reliable and easier to optimize and to control. One very 
promising possibility is to look for a temporally partially 
coherent input optical field generated by using multiple
laser beams with different wavelengths. Indeed, in a typical 
situation in which the response time of the detector is large 
compared with the inverse of the frequency difference between 
any pair of beams in the input field, rapidly oscillating 
contributions to the total intensity would average out.
As a result, one can expect smaller values of high moments of the 
intensity compared with corresponding values in the 
single wavelength case. This would in turn result in smaller values 
for the scintillation index and BER and higher values for the SNR.

We emphasize that generation of a temporally partially coherent 
source of light consisting of multiple laser beams with different wavelengths 
can be realized in a convenient and efficient manner by using 
an array of vertical external cavity surface lasers 
(VECSELs). These devices have the advantage of generating high power, high 
brightness wavelength tunable $\mbox{TEM}_{00}$  beams (lowest order 
Gaussian  beams)\cite{Fallahi2006}. Individual VECSEL devices
have been demonstrated with powers of 50 Watts running multi-mode 
\cite{PCCoherent} and up to 10 Watts, single $\mbox{TEM}_{00}$ mode and 
spectrally narrow ($<$0.1nm line width) \cite{Chilla2004}. Moreover, 
the semiconductor multiple quantum well active mirror of these devices 
should allow for multi-GHz modulation rates for data transmission.

Propagation of temporally partially coherent light in atmospheric 
turbulence was first studied by Fante \cite{Fante77,Fante79}. 
Considering a single infinite plane wave this author 
obtained approximate analytic expressions for the scintillation
index of the planar wave in the weak \cite{Fante77} and 
strong \cite{Fante79} turbulence regimes. Later studies extended 
these results to the cases of a spatially incoherent source 
\cite{Fante81} and a source that is partially coherent 
in both space and time \cite{Baykal83}. 
More recently, Kiasaleh studied the scintillation 
index for a multiwavelength infinite plane wave in 
weak atmospheric turbulence \cite{Kiasaleh2004,Kiasaleh2005}. 
This author showed that use of a 
multiwavelength plane wave leads to an increase in the 
achievable SNR as well as in the upper bound on the SNR 
imposed by increasing the aperture size. 
These previous studies focused on infinite planar waves or on spherical
waves, whereas in reality, Gaussian laser beams with finite initial 
spot size and possibly non-zero phase front radius of curvature are
employed. Since the dynamics of the optical field strongly depends on 
the initial spot size and phase front radius of curvature 
it is important to take these characteristics into account
 when calculating turbulence effects. Furthermore, the fact that 
the optical field of $N$ Gaussian laser beams depends on 
$2N$ additional parameters allows for a greater flexibility in
optimization and control of the input field against turbulence effects.
Thus, the lower values for high moments of the intensity 
combined with the greater flexibility make the setup 
based on multiple Gaussian beams with different wavelengths 
very advantageous and call for a detailed investigation of the
optical field dynamics in this setup.  
    
In this paper we take this important task and 
study propagation of two lowest order Gaussian laser beams with 
different wavelengths in weak atmospheric turbulence. 
Using the Rytov approximation and assuming 
a slow detector we calculate the longitudinal component of the
scintillation index for both the Kolmogorov spectrum and the 
Von K\'arm\'an spectrum. For a typical setup of a free space
laser communication system we show that the longitudinal 
scintillation exhibits a minimum as a function of the initial 
beam separation. We interpret this minimum as corresponding to
the optimal configuration of the two-beam system, where 
optimization is with respect to the value of the longitudinal 
scintillation. Moreover, the longitudinal 
scintillation for the optimal two-beam configuration is smaller 
by more than 50$\%$ compared with the longitudinal scintillation 
for a single lowest order Gaussian beam with the same total power.   
The existence of the longitudinal scintillation 
minimum is found to be independent of 
the turbulence spectrum. Furthermore, we calculate the total
scintillation index for the optimal two-beam configuration
using the Von K\'arm\'an spectrum and find that the values of
the total scintillation are smaller by 12$\%$-64$\%$ compared
with the corresponding values for a single Gaussian 
beam with the same total power. Further insight into this
improvement is gained by decomposing the total scintillation into 
self- and cross-intensity contributions. Finally, we show that the
radial scintillation for the optimal two-beam system is smaller
by 35$\%$-40$\%$ than the radial scintillation in the single beam case.

The rest of the paper is organized as follows. In Section 2 we present 
the method used to calculate the scintillation index as well as 
general expressions for the scintillation index of a system consisting of
$N$ Gaussian laser beams with different wavelengths. In section 3 we 
analyze in detail the total scintillation index as well as its 
longitudinal and radial components for a two-beam system considering 
a typical setup of a free space laser communication system. 
Section 4 is reserved for discussion. In Appendix A we derive 
in detail some relations appearing in Section 1.

\section{Calculation of the scintillation index}
We consider propagation of $N$ lowest order Gaussian laser beams 
with different wavelengths $\lambda_{j}$, where $j=1, ..., N$, 
in weak atmospheric turbulence. For simplicity and without loss of 
generality we assume that the beams are linearly polarized and propagate 
along the $z$ axis. We denote by ${\mathbf d_{j}}$ 
the locations of the beam centers at the input plane $z=0$.  
Thus, the magnitude of the total electric field $E$  
at $z=0$ is given by   
\begin{eqnarray}&&
E({\mathbf r},0,t)=
\sum_{j=1}^{N}E_{j}({\mathbf r_{j}},0,t)=
\sum_{j=1}^{N}U_{j}({\mathbf r_{j}},0)
\exp\left[-i\omega_{j}t\right],
\label{pcb1}
\end{eqnarray}
where 
\begin{eqnarray}&&
U_{j}({\mathbf r_{j}},0)=A_{0j}\times
\nonumber \\&&
\exp\left[-\left(\frac{1}{W_{0j}^{2}}+\frac{ik_{j}}{2F_{0j}}\right)
\left|{\mathbf r}-{\mathbf d_{j}}\right|^{2}\right].
\label{pcb2}
\end{eqnarray}
In Eqs. (\ref{pcb1}) and (\ref{pcb2}) $E_{j}$ is the electric field 
due to the $j$-th beam, ${\mathbf r}$ is the 
radius vector in the $xy$ plane, 
${\mathbf r_{j}}\equiv{\mathbf r}-{\mathbf d_{j}}$, $t$ is time,
$k_{j}=2\pi/\lambda_{j}$ are the wave numbers and
$\omega_{j}=k_{j}c$ are the angular frequencies, where $c$ is the speed of
light. In addition, $A_{0j}$ are the initial on-axis amplitudes, 
$W_{0j}$ are the initial spot sizes and $F_{0j}$ are 
the initial phase front radii of curvature. 

We assume that the intensity of the optical field is small enough so that
its evolution is governed by the linear wave equation. 
Therefore, the dynamics of the fields
$U_{j}$ is described by 
\begin{eqnarray}&&
\nabla^{2}U_{j}+k_{j}^{2}\left[1+n_{1}({\mathbf r},z)\right]^{2}U_{j}=0,
\label{linear1}
\end{eqnarray} 
where $n_{1}({\mathbf r},z)$ stands for the turbulence induced 
fluctuation in the refractive index coefficient. 
The total optical field after propagating a distance $z=L$ is    
\begin{eqnarray}&&
E({\mathbf r},L,t)=
\sum_{j=1}^{N}E_{j}({\mathbf r_{j}},L,t)=
\sum_{j=1}^{N}U_{j}({\mathbf r_{j}},L)\exp\left[-i\omega_{j}t\right].
\label{pcb3}
\end{eqnarray}
Assuming weak turbulence, $|n_{1}({\mathbf r},z)|\ll 1$, 
Eq. (\ref{linear1}) can be approximated by
\begin{eqnarray}&&
\nabla^{2}U_{j}+k_{j}^{2}\left[1+2n_{1}({\mathbf r},z)\right]U_{j}=0.
\label{linear2}
\end{eqnarray} 
To solve Eq. (\ref{linear2}) we use the Rytov perturbation method 
and express $U_{j}$ as 
\begin{eqnarray}&&
U_{j}({\mathbf r_{j}},L)=
U_{0j}({\mathbf r_{j}},L)
\exp\left[\psi_{j}(\mathbf{r_{j}},L;k_{j})\right],
\label{pcb4}
\end{eqnarray}  
where $U_{0j}$ is the field of the $j$-th beam in free space 
(in the absence of turbulence) and $\psi_{j}$ is a 
complex field describing the effects of turbulence on the $j$-th
beam and calculated perturbatively up to the second order in 
$n_{1}({\mathbf r},z)$.

Following the approach derived in Refs. \cite{Andrews98,Andrews2001}
we define the beam parameters $\Theta_{0j}$ and 
$\Lambda_{0j}$, which characterize the unperturbed beams 
in terms of the initial phase front radii of curvature 
$F_{0j}$ and spot sizes $W_{0j}$,
respectively 
\begin{eqnarray}&&
\Theta_{0j}=1-\frac{L}{F_{0j}},\;\;\;\;\;
\Lambda_{0j}=\frac{2L}{k_{j}W_{0j}^{2}}.
\label{par0}
\end{eqnarray}
We also use the beam parameters $\Theta_{j}$ and 
$\Lambda_{j}$, characterizing the unperturbed beams in terms of the
phase front radii of curvature $F_{j}$ and spot sizes $W_{j}$ at
distance $L$     
\begin{eqnarray}&&
\Theta_{j}=1-\frac{L}{F_{j}}=
\frac{\Theta_{0j}}
{\Theta_{0j}^{2}+\Lambda_{0j}^{2}},\;\;\;\;\;
\Lambda_{j}=\frac{2L}{k_{j}W_{j}^{2}}=
\frac{\Lambda_{0j}}
{\Theta_{0j}^{2}+\Lambda_{0j}^{2}}.
\label{par1}
\end{eqnarray}

The time-dependent total intensity of the $N$-beam system is 
\begin{eqnarray}&&
I({\mathbf r},L,t)=
\sum_{j=1}^{N}I_{j}({\mathbf r_{j}},L)+
\sum_{j}^{N}\sum_{m\ne j}^{N}
E_{j}({\mathbf r_{j}},L,t)
E_{m}^{*}({\mathbf r_{m}},L,t),
\label{int1}
\end{eqnarray}
where $I_{j}=|E_{j}|^{2}=|U_{j}|^{2}$ is the intensity of the $j$-th beam. 
The intensity measured by the detector can be defined 
as the following time average
\begin{eqnarray}&&
I_{det}({\mathbf r},L)\equiv
\langle I({\mathbf r},L,t)\rangle_{det}\equiv
\frac{1}{\tau}\int_{0}^{\tau}{\rm d}t I({\mathbf r},L,t),
\label{int2}   
\end{eqnarray}
where $\tau$ is the response time of the detector and 
$\langle\dots\rangle_{det}$ denotes time average in the detector.
Assuming a slow detector and different wavelengths, 
$\lambda_{j}\ne \lambda_{m}$ for $j\ne m$, we
neglect the terms $U_{j}U_{m}^{*}$
$j\ne m$, which are rapidly oscillating with time. 
[See also Ref. \cite{Kiasaleh2004}, where a similar approximation 
was made for multiple planar laser beams]. Thus, the average of the 
total intensity of the field is given by
\begin{eqnarray}&&
I_{det}({\mathbf r},L)\equiv
\langle I({\mathbf r},L,t)\rangle_{det}
\simeq
\sum_{j=1}^{N}I_{j}({\mathbf r_{j}},L).
\label{int3}
\end{eqnarray}

We are interested in calculating the total scintillation index 
$\sigma^{2}_{I}$ which is defined by
\begin{eqnarray}&&
\sigma^{2}_{I}({\mathbf r},L)\equiv
\frac{\langle I_{det}^{2}({\mathbf r},L)\rangle}
{\langle I_{det}({\mathbf r},L)\rangle^{2}}-1,
\label{pcb6}
\end{eqnarray}
where $\langle\dots\rangle$ (without any subscript) stands for
average over different realizations of turbulence disorder.
Using Eqs. (\ref{int3}) and (\ref{pcb6}) we obtain
\begin{eqnarray}&&
\sigma^{2}_{I}({\mathbf r},L)=
\frac{\sum_{j=1}^{N}\langle I_{j}^{2}({\mathbf r_{j}},L)\rangle+
2\sum_{j}^{N}\sum_{m>j}^{N} 
\langle I_{j}({\mathbf r_{j}},L)I_{m}({\mathbf r_{m}},L)\rangle}
{\left( \sum_{j=1}^{N}
\langle I_{j}({\mathbf r_{j}},L)\rangle\right)^{2}}-1.
\label{pcb6a}
\end{eqnarray}
Following the usual convention we 
decompose $\sigma^{2}_{I}$ into a longitudinal 
(${\mathbf r}$-independent) component
\begin{eqnarray}&&
\sigma^{2}_{I,l}(L)\equiv\sigma^{2}_{I}(0,L)
\label{pcb6b}
\end{eqnarray}  
and a radial component 
\begin{eqnarray}&&
\sigma^{2}_{r}({\mathbf r},L)\equiv
\sigma^{2}_{I}({\mathbf r},L)-\sigma^{2}_{I,l}(L).
\label{pcb6c}
\end{eqnarray}  

Within the framework of the Rytov approximation the 
average intensity of the $j$-th laser beam is given by \cite{Andrews98}
\begin{eqnarray}&&
\langle I_{j}({\mathbf r_{j}},L)\rangle=
I_{0j}({\mathbf r_{j}},L)
\exp\left[H_{1j}(r_{j},L)
\right],
\label{pcb7}
\end{eqnarray}
where
\begin{eqnarray}&&
I_{0j}({\mathbf r_{j}},L)=
\frac{A_{0j}^{2}W_{0j}^{2}}{W_{j}^{2}}
\exp\left[-\frac{2r_{j}^{2}}{W_{j}^{2}}\right]
\label{pcb7b}
\end{eqnarray}
and 
\begin{eqnarray}&&
H_{1j}(r_{j},L)=
4\pi^{2}k_{j}^{2}L\int_{0}^{1}{\rm d}\xi\int_{0}^{\infty}
{\rm d}\kappa\kappa\Phi_{n}(\kappa)\times
\nonumber \\&&
\left[I_{0}\left(2\Lambda_{j}r_{j}\xi\kappa\right)
\exp\left(-\frac{\Lambda_{j}L\kappa^{2}\xi^{2}}{k_{j}}\right)
-1\right].
\label{pcb7a}
\end{eqnarray}
In Eq. (\ref{pcb7a}) $\xi=1-z/L$, $\kappa$ is the wave number, 
$\Phi_{n}(\kappa)$ is the power spectral density of the 
refractive index fluctuations and $I_{0}(x)$ is the modified Bessel 
function of the first kind and zero order, $I_{0}(x)=J_{0}(ix)$, where 
$J_{0}(x)$ is the Bessel function of first kind and zero order \cite{Stegun}.
The average of the second moment 
$\langle I_{j}^{2}({\mathbf r_{j}},L)\rangle$ is given by 
\cite{Andrews98}
\begin{eqnarray}&&
\!\!\!\!\!\!
\langle I_{j}^{2}({\mathbf r_{j}},L)\rangle=
\langle I_{j}({\mathbf r_{j}},L)\rangle^{2}
\exp\left[H_{2j}(r_{j},L)\right],
\label{pcb9}
\end{eqnarray}
where
\begin{eqnarray}&&
H_{2j}(r_{j},L)=
8\pi^{2}k_{j}^{2}L\int_{0}^{1}{\rm d}\xi\int_{0}^{\infty}
{\rm d}\kappa\kappa\Phi_{n}(\kappa)
\exp\left(-\frac{\Lambda_{j}L\kappa^{2}\xi^{2}}{k_{j}}\right)
\times
\nonumber \\&&
\left\{I_{0}\left(2\Lambda_{j}r_{j}\xi\kappa\right)
-\cos\left[\frac{L\kappa^{2}\xi\left(1-\bar\Theta_{j}\xi\right)}{k_{j}}
\right]\right\}
\label{pcb11}
\end{eqnarray}
and $\bar\Theta_{j}=1-\Theta_{j}$. 
The cross-intensity term 
$\langle I_{j}({\mathbf r_{j}},L)I_{m}({\mathbf r_{m}},L)\rangle$ 
is given by (see Appendix A)
\begin{eqnarray}&&
\langle I_{j}({\mathbf r_{j}},L)I_{m}({\mathbf r_{m}},L)\rangle=
\langle I_{j}({\mathbf r_{j}},L)\rangle
\langle I_{m}({\mathbf r_{m}},L)\rangle\times
 \nonumber \\&&
\exp\left\{
E_{2jm}({\mathbf r_{j}},{\mathbf r_{m}};k_{j},k_{m})+
E_{2mj}({\mathbf r_{m}},{\mathbf r_{j}};k_{m},k_{j})+
\right.
 \nonumber \\&&
\left.
2{\mbox Re}\left[
E_{3jm}({\mathbf r_{j}},{\mathbf r_{m}};k_{j},k_{m})\right]
\right\},
\label{pcb13}
\end{eqnarray}
where     
\begin{eqnarray}&&
E_{2jm}({\mathbf r_{j}},{\mathbf r_{m}};k_{j},k_{m})=
4\pi^{2}k_{j}k_{m}L\int_{0}^{1}{\rm d}\xi\int_{0}^{\infty}
{\rm d}\kappa\kappa
\Phi_{n}(\kappa)J_{0}\left(\kappa\left|\gamma_{j}{\mathbf r_{j}}-
\gamma_{m}^{*}{\mathbf r_{m}}\right|\right)\times
 \nonumber \\&&
\exp\left[-\frac{i}{2}\kappa^{2}L
\left(\frac{\gamma_{j}}{k_{j}}-\frac{\gamma_{m}^{*}}{k_{m}}\right)\xi
\right],
\label{pcb14}
\end{eqnarray}
\begin{eqnarray}&&
E_{2mj}({\mathbf r_{m}},{\mathbf r_{j}};k_{m},k_{j})=
4\pi^{2}k_{j}k_{m}L\int_{0}^{1}{\rm d}\xi\int_{0}^{\infty}
{\rm d}\kappa\kappa
\Phi_{n}(\kappa)J_{0}\left(\kappa\left|\gamma_{m}{\mathbf r_{m}}-
\gamma_{j}^{*}{\mathbf r_{j}}\right|\right)\times
 \nonumber \\&&
\exp\left[-\frac{i}{2}\kappa^{2}L
\left(\frac{\gamma_{m}}{k_{m}}-\frac{\gamma_{j}^{*}}{k_{j}}\right)\xi
\right]
\label{pcb15}
\end{eqnarray}
and
\begin{eqnarray}&&
E_{3jm}({\mathbf r_{j}},{\mathbf r_{m}};k_{j},k_{m})=
-4\pi^{2}k_{j}k_{m}L\int_{0}^{1}{\rm d}\xi\int_{0}^{\infty}
{\rm d}\kappa\kappa
\Phi_{n}(\kappa)J_{0}\left(\kappa\left|\gamma_{j}{\mathbf r_{j}}-
\gamma_{m}{\mathbf r_{m}}\right|\right)\times
 \nonumber \\&&
\exp\left[-\frac{i}{2}\kappa^{2}L
\left(\frac{\gamma_{j}}{k_{j}}+\frac{\gamma_{m}}{k_{m}}\right)\xi
\right].
\label{pcb16}
\end{eqnarray}
In Eqs. (\ref{pcb14}-\ref{pcb16})  
$\gamma_{j}=1-(\bar\Theta_{j}+i\Lambda_{j})\xi$.

\section{Specific setups, optimal configuration and reduction
of the total scintillation index}
We consider in detail a system with two lowest order 
Gaussian beams whose wavelengths are
$\lambda_{1}=1.0\times 10^{-6}$m and $\lambda_{2}=1.01\times 10^{-6}$m,
and whose initial spot sizes are $W_{01}=W_{02}=1.0$cm.
The initial locations of the spot centers are 
${\mathbf d_{1}}=-d{\mathbf {\hat y}}/2$ and 
${\mathbf d_{2}}=d{\mathbf {\hat y}}/2$ so that the initial beam separation
is $d$. We assume that the initial on-axis 
amplitudes are equal $A_{01}=A_{02}$
and that the beams are initially collimated so that $F_{01}=F_{02}=\infty$.
We use the Von K\'arm\'an spectrum to describe the refractive 
index fluctuations
\begin{eqnarray}&&
\Phi_{n}(\kappa)=0.033C_{n}^{2}
\frac{\exp\left(-\kappa^{2}/\kappa_{in}^{2}\right)}
{\left(\kappa^{2}+\kappa_{out}^{2}\right)^{11/6}}
\label{pcb18}
\end{eqnarray}
where $\kappa_{in}=5.92/l_{0}$, $\kappa_{out}=1/L_{0}$, 
$l_{0}$ and $L_{0}$ are the turbulence inner and outer scales, respectively,
and $C_{n}^{2}$ is the refractive index structure parameter.
We choose $C_{n}^{2}=3.0\times 10^{-15}\mbox{m}^{-2/3}$, corresponding to 
weak atmospheric turbulence conditions, and $l_{0}=1.0$mm, $L_{0}$=1.0m.
We concentrate on the statistics at a propagation distance $L=1$km, 
where the Rytov variance
\begin{eqnarray}&&
\sigma_{R}^{2}=1.23C_{n}^{2}k^{7/6}L^{11/6}
\label{pcb19}
\end{eqnarray}
is about $0.1$ for both beams and one can indeed employ the
Rytov perturbation method. The free space spot sizes of the 
beams at this distance are about 3.3cm each.   
 
\begin{figure}[ptb]
\epsfxsize=12.0cm \epsffile{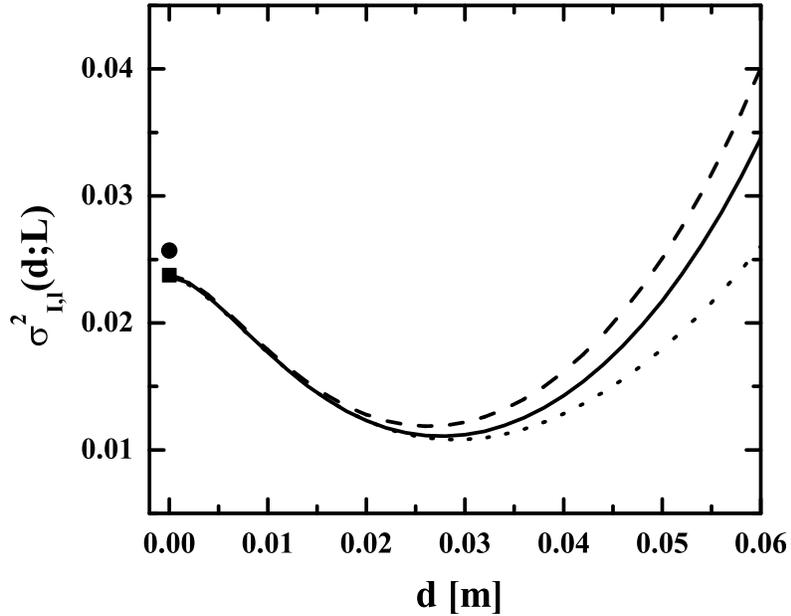} 
\caption{Longitudinal component of the scintillation index 
$\sigma^{2}_{I,l}$ as a function of the initial beam separation $d$.
The solid line is the result obtained by using the
the Von K\'arm\'an spectrum. The dashed line is the result obtained by
using the Kolmogorov spectrum and the dotted line corresponds to
the result obtained by using the Kolmogorov spectrum and assuming
that $r_{j}/W_{j}\ll 1$. The square/circle stand for the 
longitudinal scintillation of a single beam with the same total power and the
same initial spot size/amplitude, respectively.}
 \label{fig1}
\end{figure}

Using the above relations we first calculate the 
longitudinal component of the scintillation index as 
defined by Eq. (\ref{pcb6b}). The longitudinal component is of 
special importance when the spot size is large compared with 
the radius of the receiver's collecting lens since 
in this case the average signal to noise ratio is 
predominantly determined by the on-axis values of 
$\langle I\rangle$ and $\langle I^{2}\rangle$. The dependence of 
$\sigma^{2}_{I,l}$ on the initial separation between the two beams $d$ 
for the aforementioned values of the parameters is
plotted in Fig. \ref{fig1} (solid line). 
It can be seen that the curve $\sigma^{2}_{I,l}(d;L)$ possesses a 
minimum at an intermediate value of $d$, $d_{0}=2.8$cm. Clearly 
this minimum corresponds to the optimal configuration of the two-beam system
for the given physical parameters, where optimization is 
performed with respect to longitudinal scintillation. 
The figure also shows a comparison with the longitudinal scintillation of a 
single lowest order Gaussian beam with the same total power 
$P_{0}=\pi A_{0}^{2}W_{0}^{2}$ for the following two cases: 
(a) the single beam has the same initial spot size as that of each of the
two beams (square); (b) the single beam has the same amplitude as 
that of each of the two beams (circle). 
One can see that the two-beam system gives a 53.4$\%$ reduction
of the longitudinal scintillation in comparison with the single beam value 
in case (a) and a 56.9$\%$  reduction in comparison with the single 
beam result in case (b). The dashed line in Fig. \ref{fig1} corresponds to 
the result obtained by using the Kolmogorov spectrum
\begin{eqnarray}&&
\Phi_{n}(\kappa)=0.033C_{n}^{2}\kappa^{-11/3}.
\label{pcb18a}
\end{eqnarray}   
The dotted line corresponds to the approximate result obtained 
by using the Kolmogorov spectrum (\ref{pcb18a}), assuming that for each beam 
$r_{j}<W_{j}$, and expanding up to the second order with respect to
$r_{j}/W_{j}$. The comparison of these two results with the result based
on the Von K\'arm\'an spectrum shows that for the parameters considered 
here the existence of the minimum of $\sigma^{2}_{I,l}(d;L)$ 
is not very sensitive to the details of the spectrum model.
In addition, the minimum value of the longitudinal scintillation 
and the location of the minimum $d_{0}$ obtained by using the more realistic 
Von K\'arm\'an spectrum are slightly smaller than the values obtained by 
using the Kolmogorov spectrum.

\begin{figure}[ptb]
\epsfxsize=12.0cm \epsffile{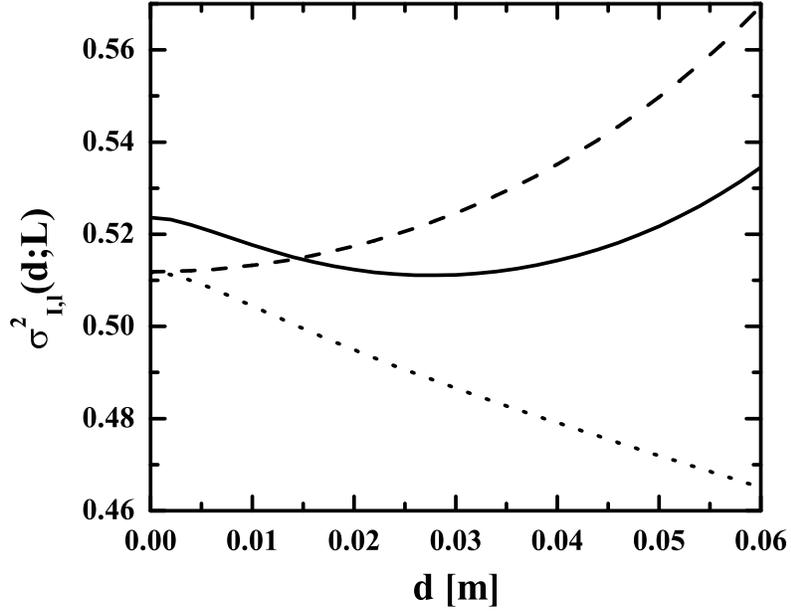} 
\caption{Self- and cross-intensity contributions to the longitudinal
scintillation index $\sigma^{2}_{I,l,s}$ and $\sigma^{2}_{I,l,c}$,
respectively, vs beam separation $d$. The dashed line represents
$\sigma^{2}_{I,l,s}(d;L)$ and the dotted line stands for 
$\sigma^{2}_{I,l,c}(d;L)$. The solid line corresponds to 
$\sigma^{2}_{I,l}(d;L)+1/2$.}
 \label{fig2}
\end{figure}

In order to better understand the origin 
of the minimum of the longitudinal component of the scintillation
index we further decompose $\sigma^{2}_{I,l}$  into a self-intensity 
contribution $\sigma^{2}_{I,l,s}$ and a cross-intensity contribution 
$\sigma^{2}_{I,l,c}$ in the following manner
\begin{eqnarray}&&
\sigma^{2}_{I,l}=\sigma^{2}_{I,l,s}+\sigma^{2}_{I,l,c}-1,
\label{pcb20}
\end{eqnarray}
where
\begin{eqnarray}&&
\sigma^{2}_{I,l,s}(L)\equiv
\frac{\langle I_{1}^{2}(d{\mathbf {\hat y}}/2,L)\rangle+
\langle I_{2}^{2}(-d{\mathbf {\hat y}}/2,L)\rangle}
{\langle I({\mathbf 0},L)\rangle^{2}}
\label{pcb21}
\end{eqnarray}
and
\begin{eqnarray}&&
\sigma^{2}_{I,l,c}(L)\equiv
\frac{2\langle I_{1}(d{\mathbf {\hat y}}/2,L)
I_{2}(-d{\mathbf {\hat y}}/2,L)\rangle}
{\langle I({\mathbf 0},L)\rangle^{2}}.
\label{pcb22}
\end{eqnarray}
The $d$ dependence of the two components 
$\sigma^{2}_{I,l,s}$ and $\sigma^{2}_{I,l,c}$ is shown in Fig. \ref{fig2}
together with the $d$ dependence of the total $\sigma^{2}_{I,l}$. 
As expected, the cross-intensity component 
$\sigma^{2}_{I,l,c}$ is a decreasing function
of $d$. In contrast, $\sigma^{2}_{I,l,s}$ is increasing 
with increasing $d$ due to the radial scintillation contribution 
of each beam. Thus, the existence of a minimum for 
$\sigma^{2}_{I,l}(d;L)$ is a direct 
consequence of this opposite behavior of the self- and cross-intensity 
contributions with increasing beam separation.

\begin{figure}[ptb]
\epsfxsize=12.0cm \epsffile{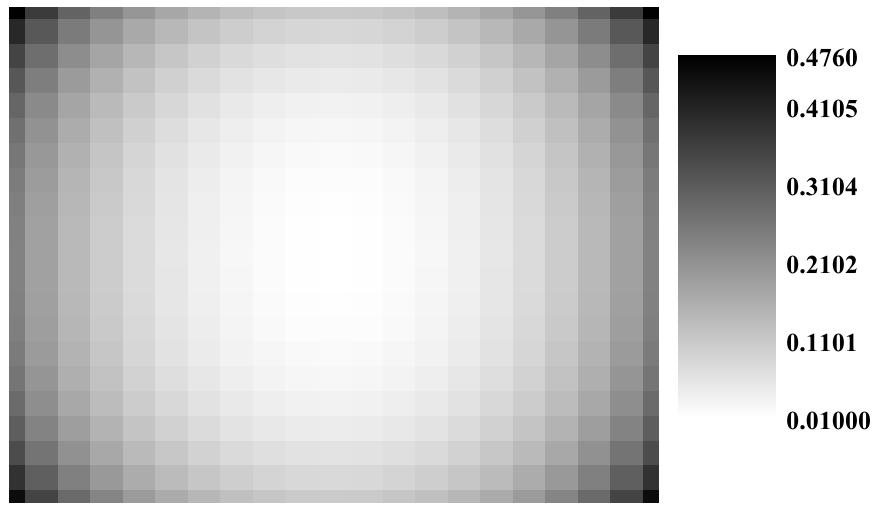} 
\caption{Total scintillation index  $\sigma^{2}_{I}({\mathbf r},L)$ 
for the two-beam system with the optimal configuration $d_{0}=2.8$cm at 
a propagation distance $L=1$km. The figure shows a 8cm$\times$8cm 
domain centered about the $z$ axis.}
 \label{fig3}
\end{figure}

When the spot size is comparable with the radius 
of the receiver's collecting lens the radial 
component of the scintillation index becomes important and one
should take into account both longitudinal and radial contributions. 
In Figure \ref{fig3} we present the total scintillation index of the 
two-beam system for the optimal configuration, i.e. for $d=d_{0}=2.8$cm,  
in a 8cm$\times$8cm domain centered about the $z$ axis. 
The total scintillation index attains values in the range
$0.011<\sigma^{2}_{I}<0.474$. The minimum 
value of the scintillation is attained on the $z$ axis 
since the two beams are already strongly overlapping at this 
distance, i.e., $d_{0}/2<W_{1}, W_{2}$.

\begin{figure}[ptb]
\epsfxsize=12.0cm \epsffile{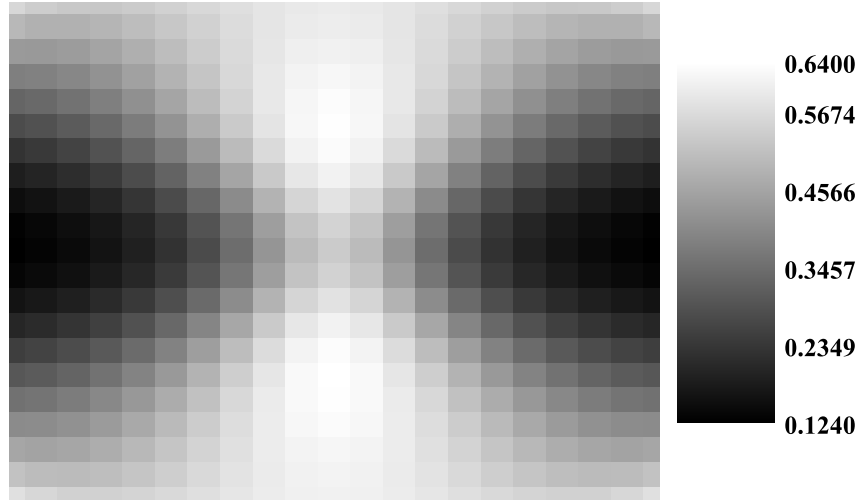} 
\caption{Fractional reduction of the total scintillation index 
$R_{a}({\mathbf r},L)$ obtained by using the optimal two-beam system 
relative to a single Gaussian beam with the same total 
intensity and initial spot size.}
 \label{fig4}
\end{figure}

It is very instructive to analyze the reduction in the total 
scintillation index obtained by using the optimal two-beam system
relative to a single lowest order Gaussian 
beam for cases (a) and (b) mentioned above. 
For this purpose we denote by $\sigma^{2}_{I,a}$ and 
$\sigma^{2}_{I,b}$ the total scintillation of a single beam 
in cases (a) and (b), respectively,  
and define the fractional reduction in the total
scintillation relative to the single beam 
values in the two cases as
\begin{eqnarray}&&
R_{a}({\mathbf r},L)=\frac{\sigma^{2}_{I,a}({\mathbf r},L)-
\sigma^{2}_{I}({\mathbf r},L)}
{\sigma^{2}_{I,a}({\mathbf r},L)}
\label{pcb23a}
\end{eqnarray}   
and
\begin{eqnarray}&&
R_{b}({\mathbf r},L)=\frac{\sigma^{2}_{I,b}({\mathbf r},L)-
\sigma^{2}_{I}({\mathbf r},L)}
{\sigma^{2}_{I,b}({\mathbf r},L)}.
\label{pcb23b}
\end{eqnarray}
Figure \ref{fig4} shows $R_{a}({\mathbf r},L)$ 
in the 8cm$\times$8cm domain.   
It can be seen that in case (a) the fractional reduction factor 
is in the range $0.124<R_{a}<0.639$ which means that the
reduction is larger than 12.4$\%$ everywhere within the 
8cm$\times$8cm domain and can be as 
large as 63.9$\%$. Similar calculation shows that the relative reduction in 
the total scintillation in case (b) is even larger. 
In both cases, the largest improvement is obtained along the
$y$ axis and in the corners of the domain. The smallest improvement 
is along the $x$ axis, where the total intensity is initially
small. This behavior suggests that in a 4-beam system further increase
of the value of $R$ can be obtained by locating the 
two additional beams along the $x$ axis.        

\begin{figure}[ptb]
\epsfxsize=12.0cm \epsffile{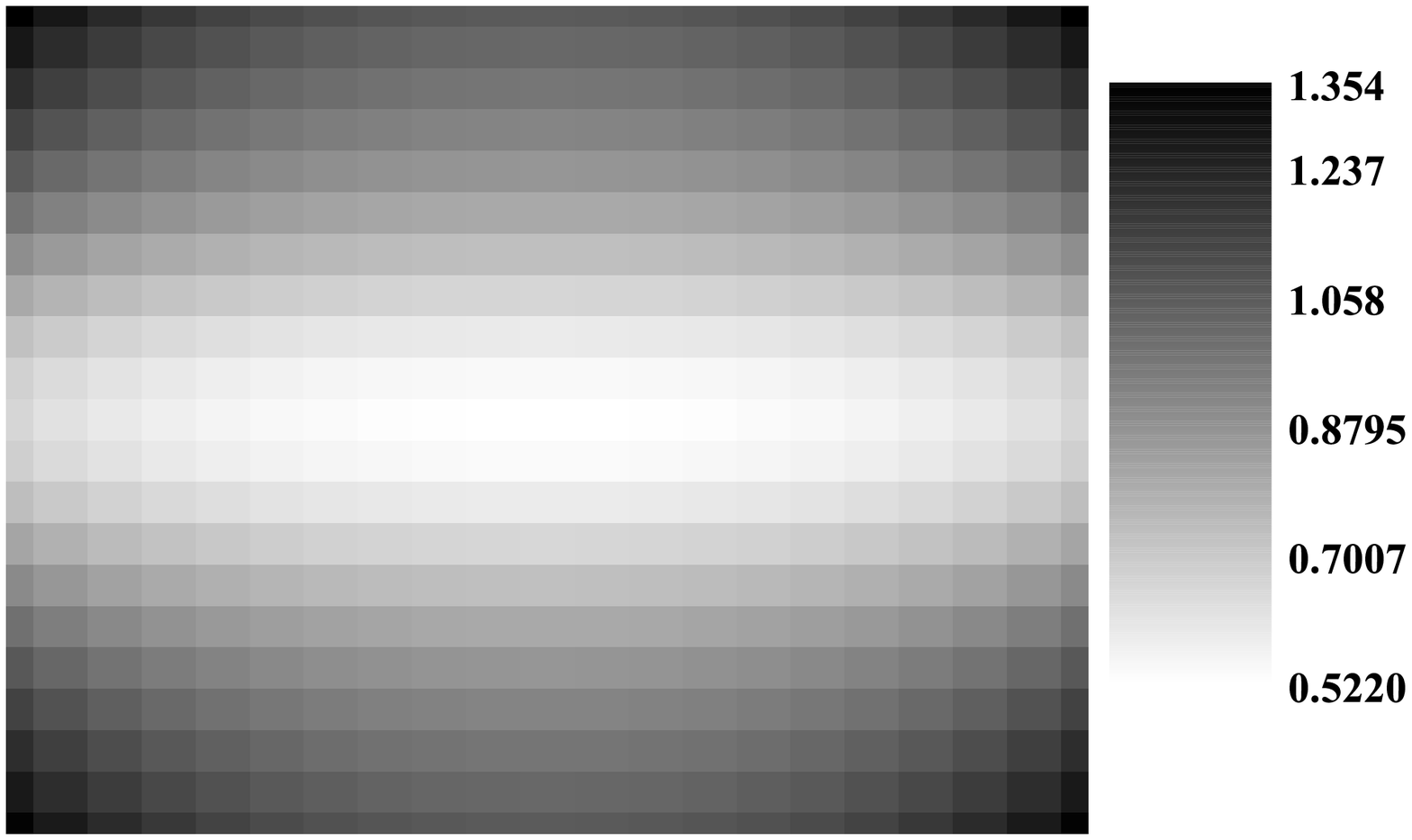} 
\caption{Self-intensity contribution to the total scintillation index
$\sigma^{2}_{I,s}({\mathbf r},L)$ for the optimal two-beam system.}
 \label{fig6}
\end{figure}

\begin{figure}[ptb]
\epsfxsize=12.0cm \epsffile{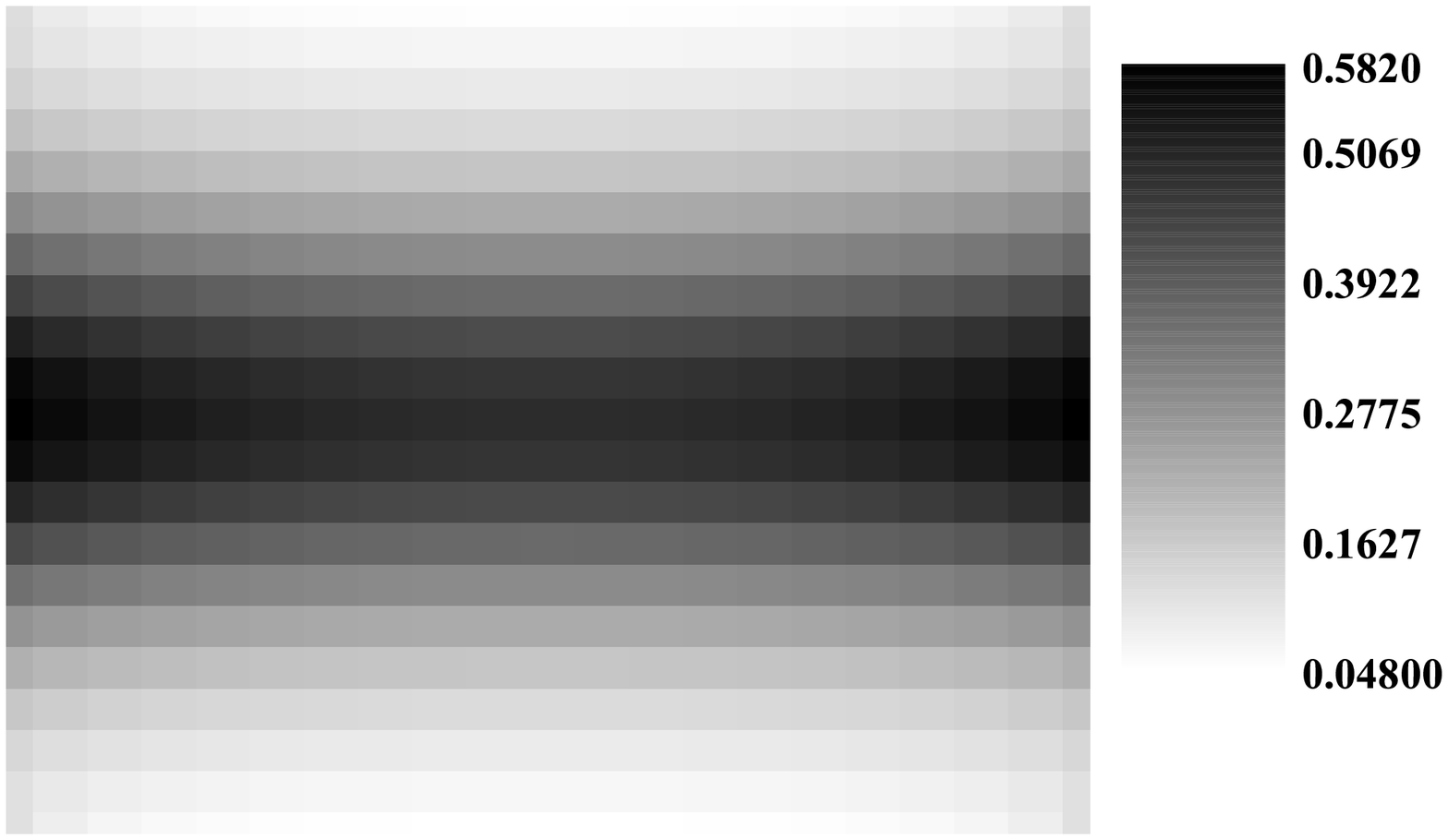} 
\caption{Cross-intensity contribution to the total scintillation index
$\sigma^{2}_{I,c}({\mathbf r},L)$ for the optimal two-beam system.}
 \label{fig7}
\end{figure}

To get further insight into the results presented in Figs. \ref{fig3}
and \ref{fig4} we decompose the total scintillation index 
into self- and cross-intensity contributions 
$\sigma^{2}_{I,s}$ and $\sigma^{2}_{I,c}$ in the following manner
\begin{eqnarray}&&
\sigma^{2}_{I}({\mathbf r},L)=\sigma^{2}_{I,s}({\mathbf r},L)+
\sigma^{2}_{I,c}({\mathbf r},L)-1,
\label{pcb24}
\end{eqnarray}
where
\begin{eqnarray}&&
\sigma^{2}_{I,s}({\mathbf r},L)\equiv
\frac{\langle I_{1}^{2}({\mathbf r_{1}},L)\rangle+
\langle I_{2}^{2}({\mathbf r_{2}},L)\rangle}
{\langle I({\mathbf r},L)\rangle^{2}}
\label{pcb25}
\end{eqnarray}
and
\begin{eqnarray}&&
\sigma^{2}_{I,c}({\mathbf r},L)\equiv
\frac{2\langle I_{1}({\mathbf r_{1}},L)
I_{2}({\mathbf r_{2}},L)\rangle}
{\langle I({\mathbf r},L)\rangle^{2}}.
\label{pcb26}
\end{eqnarray}
The self- and cross-intensity contributions to the total scintillation index 
for the optimal two-beam system are shown in Figures \ref{fig6} and \ref{fig7},
respectively. The self-intensity contribution attains values 
in the range $0.523<\sigma^{2}_{I,s}<1.354$ whereas the values of the 
cross-intensity contribution are in the range $0.052<\sigma^{2}_{I,c}<0.582$.
The largest values of $\sigma^{2}_{I,s}$ are attained in the corners of the 
8cm$\times$8cm domain, where the $\sigma^{2}_{I,c}$ contribution 
is negligible.  It is also seen that
the values of $\sigma^{2}_{I,s}$ are larger along the $y$ axis and smaller
along the $x$ axis. This behavior is due to the contributions coming from 
the radial scintillation of each beam. The largest $\sigma^{2}_{I,c}$ 
values are attained on the $x$ axis. These values are comparable 
to the values of $\sigma^{2}_{I,s}$ on the $x$ axis which 
explains the relatively small values of $R_{a}$ seen along 
the $x$ axis in Fig. \ref{fig4}.

\begin{figure}[ptb]
\epsfxsize=12.0cm \epsffile{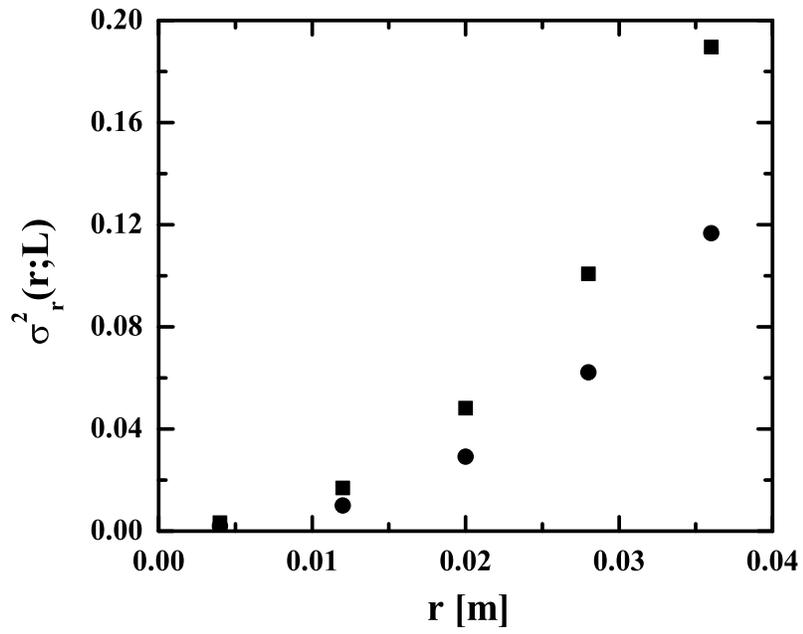} 
\caption{Circularly averaged radial scintillation index for the
optimal two-beam system $\sigma^{2}_{rr}$ as a function of radius $r$ 
(circles). The squares correspond to the result obtained for a single 
beam with the same power and initial spot size.}
 \label{fig8}
\end{figure}

In analyzing the behavior of the radial scintillation we first note
that unlike the situation in the single-beam case, in the two-beam
case $\sigma^{2}_{r}({\mathbf r},L)$ is not radially symmetric. 
To enable comparison with the single-beam result we define 
the circularly averaged radial scintillation 
in the two-beam case $\sigma^{2}_{rr}$ as the average 
over angle $\theta$ of the radial scintillation index: 
$\sigma^{2}_{rr}(r,L)\equiv 
\langle\sigma^{2}_{r}({\mathbf r},L)\rangle_{\theta}$. 
The $r$-dependence of $\sigma^{2}_{rr}$ for the optimal two-beam
system is shown in Fig. \ref{fig8}.  The figure also shows a
comparison with the radial scintillation 
of a single lowest order Gaussian beam with 
the same total power and initial spot size. It can be seen that the
value of the radial scintillation index for the optimal 
two-beam system at a given $r$ is smaller by 
about 35$\%$-40$\%$ than the corresponding value for the single beam.
Therefore, optimizing the two-beam system with respect to the 
longitudinal scintillation also leads to a significant reduction in the
radial scintillation.   

\section{Conclusions}
We investigated the dynamics of two lowest order 
Gaussian laser beams with different 
wavelengths in weak atmospheric turbulence. Assuming a 
Von K\'arm\'an turbulence spectrum and slow detector response and
using the Rytov approximation we calculated the longitudinal and
radial components of the scintillation index for a typical free
space laser communication setup. We found that the longitudinal 
scintillation possesses a minimum as a function of the initial beam
separation. This minimum corresponds to the ideal configuration of the
two beams, where optimization is performed with respect to the 
longitudinal scintillation index. The longitudinal scintillation for
the optimal two-beam configuration is smaller by more than 50$\%$ 
compared with the value for a single 
lowest order Gaussian beam with the same total power. 
We introduced the self- and cross-intensity contributions 
to the longitudinal scintillation index and explained the existence 
of the minimum in terms of the opposite behavior of these contributions
with increasing beam separation. Similar calculations of the 
longitudinal scintillation with the Kolmogorov spectrum 
show that the existence of the minimum is not very sensitive 
to the form of the turbulence spectrum. 

In actual applications the radial component of the scintillation
index might be as important as the longitudinal component. We 
therefore calculated the total scintillation index for the optimal 
two-beam configuration. We found that for the same typical setup 
considered above the values of the total scintillation are smaller by 
12$\%$-64$\%$ compared with the values of the 
total scintillation for a single beam with the same total power. 
Further analysis showed that the reduction of the scintillation 
obtained by using the optimal two-beam configuration was largest along
the $y$ axis, where the centers of the two beams are initially located, 
and smallest along the $x$ axis, where the total intensity is initially
small. This behavior is attributed to the relatively large values 
of the cross-intensity contributions to the total scintillation 
along the $x$ axis. It also suggests that in a four-beam configuration the 
system's performance can be further improved by locating 
the two additional beams along the $x$-axis. Finally, we showed  
that the circularly averaged radial scintillation index
for the optimal two-beam configuration is smaller by 
35$\%$-40$\%$ compared with the radial scintillation 
for a single beam with the same total power.   
      
The results presented in this paper open many promising 
pathways for future theoretical research. A natural extension of this
study is to investigate a system with $N>2$ Gaussian beams and
to characterize the $N$-dependence of the scintillation reduction
relative to the single beam case. Another possible direction is 
to study the dependence of the scintillation index in the
multi-beam multi-wavelength case on the convergence/divergence
of the beams, i.e., on the values of the phase front radii of curvature.
Optimization with respect to the radial 
scintillation index at a given radius is 
yet another interesting problem. From the applications point of view
it would be very interesting to evaluate the average signal to noise
ratio and bit error rate of the multi-beam multi-wavelength system.
This will require calculation of the optical field after the 
receiver's collecting lens as well as calculation of 
aperture averaging effects. Such calculations can be 
carried out in a straightforward manner by employing the 
ABCD ray matrix theory \cite{Andrews98,Andrews2001}.

\section*{Acknowledgments}
We thank M. Kolesik, E. M. Wright, A. Marathay, J. T. Murray, 
P. Polynkin, M. Mansuripur and G. Gbur for very useful discussions. 
This research was sponsored by 
the Air Force Office for Scientific Research, Air Force Material 
Command, USAF, under grant AFOSR FA9550-04-1-0213.

\appendix
\section{Derivation of Eqs. (\ref{pcb13})-(\ref{pcb16})}
\label{derivation}
In this appendix we derive Eqs. (\ref{pcb13})-(\ref{pcb16}) for the 
cross-intensity term 
$\langle I_{j}({\mathbf r_{j}},L)I_{m}({\mathbf r_{m}},L)\rangle$. 
The derivation is based on the second order Rytov approximation and 
follows the outline of the calculation of 
$\langle I^{2}({\mathbf r},L)\rangle$ for a single lowest order Gaussian 
laser beam presented in chapters 5 and 6 of Ref. \cite{Andrews98}. 
Within the framework of the Rytov approximation 
the optical field of the $j$-th beam is 
given by Eq. (\ref{pcb4}). Therefore, the intensity of the $j$-th
beam is 
\begin{eqnarray}&&
I_{j}({\mathbf r_{j}},L)=
I_{0j}({\mathbf r_{j}},L)
\exp\left[\psi_{j}(\mathbf{r_{j}},L;k_{j})+
\psi_{j}^{*}(\mathbf{r_{j}},L;k_{j})
\right],
\label{app1}
\end{eqnarray}  
where $I_{0j}({\mathbf r_{j}},L)$ is 
given by Eq. (\ref{pcb7b}). It follows that
\begin{eqnarray}&&
\langle I_{j}({\mathbf r_{j}},L)I_{m}({\mathbf r_{m}},L)\rangle=
I_{0j}({\mathbf r_{j}},L)I_{0m}({\mathbf r_{m}},L)
\exp\left[\Psi_{jm}^{(tot)}(\mathbf{r},L)
\right],
\label{app2}
\end{eqnarray}  
where
\begin{eqnarray}&&
\Psi_{jm}^{(tot)}(\mathbf{r},L)=
\psi_{j}(\mathbf{r_{j}},L;k_{j})+
\psi_{j}^{*}(\mathbf{r_{j}},L;k_{j})+
\psi_{m}(\mathbf{r_{m}},L;k_{m})+
\psi_{m}^{*}(\mathbf{r_{m}},L;k_{m}).
\label{app3}
\end{eqnarray}
Assuming that $\Psi_{ij}^{(tot)}$ is a Gaussian random variable 
\cite{Andrews98} we can use the relation
\begin{eqnarray}&&
\langle \exp\left[\Psi_{jm}^{(tot)}(\mathbf{r},L)
\right]\rangle=
\exp\left\{
\langle\Psi_{jm}^{(tot)}(\mathbf{r},L)\rangle +
\left[\langle\Psi_{jm}^{(tot)2}(\mathbf{r},L)\rangle-
\langle\Psi_{jm}^{(tot)}(\mathbf{r},L)\rangle^{2}\right]
\right\}.
\label{app4}
\end{eqnarray}  
We expand $\Psi_{jm}^{(tot)}(\mathbf{r},L)$ up to the second order 
with respect to $n_{1}({\mathbf r},z)$
\begin{eqnarray}&&
\Psi_{jm}^{(tot)}(\mathbf{r},L)\simeq
\Psi_{jm1}^{(tot)}(\mathbf{r},L)+
\Psi_{jm2}^{(tot)}(\mathbf{r},L)\simeq
 \nonumber \\&&
\left[\psi_{j1}(\mathbf{r_{j}},L;k_{j})+
\psi_{j1}^{*}(\mathbf{r_{j}},L;k_{j})+
\psi_{m1}(\mathbf{r_{m}},L;k_{m})+
\psi_{m1}^{*}(\mathbf{r_{m}},L;k_{m})
\right]+
 \nonumber \\&&
\left[\psi_{j2}(\mathbf{r_{j}},L;k_{j})+
\psi_{j2}^{*}(\mathbf{r_{j}},L;k_{j})+
\psi_{m2}(\mathbf{r_{m}},L;k_{m})+
\psi_{m2}^{*}(\mathbf{r_{m}},L;k_{m})
\right],
\label{app5}
\end{eqnarray} 
where the subscripts 1 and 2 in $\psi_{1}$ and $\psi_{2}$ denote 
first and second order. 
Since $\langle n_{1}({\mathbf r},z)\rangle=0$
\begin{eqnarray}&&
\langle \Psi_{jm}^{(tot)}(\mathbf{r},L)\rangle\simeq
\langle \Psi_{jm2}^{(tot)}(\mathbf{r},L)\rangle=
\langle\psi_{j2}(\mathbf{r_{j}},L;k_{j})\rangle+
\langle\psi_{j2}^{*}(\mathbf{r_{j}},L;k_{j})\rangle+
 \nonumber \\&&
\langle\psi_{m2}(\mathbf{r_{m}},L;k_{m})\rangle+
\langle\psi_{m2}^{*}(\mathbf{r_{m}},L;k_{m})\rangle
\label{app6}
\end{eqnarray}
and 
\begin{eqnarray}&&
\langle \Psi_{jm}^{(tot)}(\mathbf{r},L)\rangle^{2}\simeq
\langle \Psi_{jm1}^{(tot)}(\mathbf{r},L)\rangle^{2}=0.
\label{app7}
\end{eqnarray}  
In addition,
\begin{eqnarray}&&
\langle \Psi_{jm}^{(tot)2}(\mathbf{r},L)\rangle\simeq
\langle \Psi_{jm1}^{(tot)2}(\mathbf{r},L)\rangle=
\langle\psi_{j1}^{2}(\mathbf{r_{j}},L;k_{j})\rangle+
\langle\psi_{j1}^{*2}(\mathbf{r_{j}},L;k_{j})\rangle+
 \nonumber \\&&
\langle\psi_{m1}^{2}(\mathbf{r_{m}},L;k_{m})\rangle+
\langle\psi_{m1}^{*2}(\mathbf{r_{m}},L;k_{m})\rangle+
E_{2jj}({\mathbf r_{j}},{\mathbf r_{j}};k_{j},k_{j})+
 \nonumber \\&&
E_{2mm}({\mathbf r_{m}},{\mathbf r_{m}};k_{m},k_{m})+
E_{2jm}({\mathbf r_{j}},{\mathbf r_{m}};k_{j},k_{m})+
E_{2mj}({\mathbf r_{m}},{\mathbf r_{j}};k_{m},k_{j})+
 \nonumber \\&&
2{\mbox Re}\left[
E_{3jm}({\mathbf r_{j}},{\mathbf r_{m}};k_{j},k_{m})\right],
\label{app8}
\end{eqnarray} 
where
\begin{eqnarray}&&
E_{2jm}({\mathbf r_{j}},{\mathbf r_{m}};k_{j},k_{m})=
\langle\psi_{j1}(\mathbf{r_{j}},L;k_{j})
\psi_{m1}^{*}(\mathbf{r_{m}},L;k_{m})\rangle
\label{app9}
\end{eqnarray} 
and 
\begin{eqnarray}&&
E_{3jm}({\mathbf r_{j}},{\mathbf r_{m}};k_{j},k_{m})=
\langle\psi_{j1}(\mathbf{r_{j}},L;k_{j})
\psi_{m1}(\mathbf{r_{m}},L;k_{m})\rangle.
\label{app10}
\end{eqnarray}     
Denoting
\begin{eqnarray}&&
E_{1j}({\mathbf r_{j}};k_{j})=
\langle\psi_{j2}(\mathbf{r_{j}},L;k_{j})\rangle+
\frac{1}{2}\langle\psi_{j1}^{2}(\mathbf{r_{j}},L;k_{j})\rangle
\label{app11}
\end{eqnarray} 
and using Eqs. (\ref{app4})-(\ref{app8}) we obtain
\begin{eqnarray}&&
\langle \exp\left[\Psi_{ij}^{(tot)}(\mathbf{r},L)
\right]\rangle=
\exp\left\{
2E_{1j}({\mathbf r_{j}};k_{j})+
2E_{1m}({\mathbf r_{m}};k_{m})+
\right.
 \nonumber \\&&
\left.
E_{2jj}({\mathbf r_{j}},{\mathbf r_{j}};k_{j},k_{j})+
E_{2mm}({\mathbf r_{m}},{\mathbf r_{m}};k_{m},k_{m})+
E_{2jm}({\mathbf r_{j}},{\mathbf r_{m}};k_{j},k_{m})+
\right.
 \nonumber \\&&
\left.
E_{2mj}({\mathbf r_{m}},{\mathbf r_{j}};k_{m},k_{j})+
2{\mbox Re}\left[
E_{3jm}({\mathbf r_{j}},{\mathbf r_{m}};k_{j},k_{m})\right]
\right\}.
\label{app12}
\end{eqnarray}
Noting that 
\begin{eqnarray}&&
\langle I_{j}({\mathbf r_{j}},L)\rangle=
I_{0j}({\mathbf r_{j}},L)
\exp\left[
2E_{1j}({\mathbf r_{j}};k_{j})+
E_{2jj}({\mathbf r_{j}},{\mathbf r_{j}};k_{j},k_{j})
\right],
\label{app13}
\end{eqnarray}
(see also Ref. \cite{Andrews98} p. 130) we arrive at
\begin{eqnarray}&&
\langle I_{j}({\mathbf r_{j}},L)I_{m}({\mathbf r_{m}},L)\rangle=
\langle I_{j}({\mathbf r_{j}},L)\rangle
\langle I_{m}({\mathbf r_{m}},L)\rangle\times
 \nonumber \\&&
\exp\left\{
E_{2jm}({\mathbf r_{j}},{\mathbf r_{m}};k_{j},k_{m})+
E_{2mj}({\mathbf r_{m}},{\mathbf r_{j}};k_{m},k_{j})+
\right.
 \nonumber \\&&
\left.
2{\mbox Re}\left[
E_{3jm}({\mathbf r_{j}},{\mathbf r_{m}};k_{j},k_{m})\right]
\right\},
\label{app14}
\end{eqnarray}
which is Eq. (\ref{pcb13}).
   
Next we derive Eq. (\ref{pcb14}) for 
$E_{2jm}({\mathbf r_{j}},{\mathbf r_{m}};k_{j},k_{m})$. 
Within the framework of the Rytov perturbation method 
the first order term $\psi_{j1}(\mathbf{r_{j}},L;k_{j})$ 
is given by (see Ref. \cite{Andrews98} p. 105)
\begin{eqnarray}&&
\psi_{j1}(\mathbf{r_{j}},L;k_{j})=
ik_{j}\int_{0}^{L}{\rm d}z
\int\!\!\!\int{\rm d}\nu(\mathbf{K},z)
\exp\left[i\gamma_{j}(z)\mathbf{K}\cdotp\mathbf{r_{j}}-
\frac{i\kappa^{2}\gamma_{j}(z)(L-z)}{2k_{j}}
\right],
\label{app21}
\end{eqnarray}
where
\begin{eqnarray}&&
\gamma_{j}(z)=\frac{1+\alpha_{j} z}{1+\alpha_{j} L}
\label{app22}
\end{eqnarray}
and
\begin{eqnarray}&&
\alpha_{j}=
\frac{2}{k_{j}W_{0j}^{2}}+
\frac{i}{F_{0j}}.
\label{app23}
\end{eqnarray}
In Eq. (\ref{app21}) $\nu(\mathbf{K},z)$ is the random amplitude of
the refractive index fluctuations 
\begin{eqnarray}&&
n_{1}(\mathbf{r},z)=
\int\!\!\!\int{\rm d}\nu(\mathbf{K},z)
\exp\left(i\mathbf{K}\cdotp\mathbf{r}\right),
\label{app24}
\end{eqnarray}
$\mathbf{K}=(\kappa_{x},\kappa_{y})$ is the 
two-dimensional wave vector and 
$\kappa=(\kappa_{x}^{2}+\kappa_{y}^{2})^{1/2}$.
Substituting Eq. (\ref{app21}) into Eq. (\ref{app9}) and using the
correlation relation
\begin{eqnarray}&&
\langle \nu(\mathbf{K},z)\nu^{*}(\mathbf{K'},z')\rangle=
F_{n}(\mathbf{K},|z-z'|)\delta(\mathbf{K}-\mathbf{K'})
{\rm d}^{2}\mathbf{K}{\rm d}^{2}\mathbf{K'}
\label{app25}
\end{eqnarray}
where $\delta(\mathbf{K})$ stands for the Dirac delta function 
and $F_{n}(\mathbf{K},|z-z'|)$ is the two-dimensional spectral
density of the refractive index fluctuations we obtain
\begin{eqnarray}&&
E_{2jm}({\mathbf r_{j}},{\mathbf r_{m}};k_{j},k_{m})=
k_{j}k_{m}\int_{0}^{L}{\rm d}z\int_{0}^{L}{\rm d}z'
\int\!\!\!\int{\rm d}^{2}\mathbf{K}
F_{n}(\mathbf{K},|z-z'|)\times
 \nonumber \\&&
\exp\left\{i\mathbf{K}\cdotp
\left[\gamma_{j}(z)\mathbf{r_{j}}-
\gamma_{m}^{*}(z')\mathbf{r_{m}}\right]
-\frac{i\kappa^{2}}{2}
\left[\frac{\gamma_{j}(z)(L-z)}{k_{j}}-
\frac{\gamma_{m}^{*}(z')(L-z')}{k_{m}}
\right]
\right\}.
\label{app26}
\end{eqnarray}
We now change variables from $z$ and $z'$ to $\mu=z-z'$ and 
$\eta=(z+z')/2$. In addition, we use the fact that 
$F_{n}(\mathbf{K},|\mu|)$ is centered about $\mu=0$ to extend
the integration over $\mu$ to $\pm\infty$ and to take
$z=z'=\eta$. This calculation yields   
\begin{eqnarray}&&
E_{2jm}({\mathbf r_{j}},{\mathbf r_{m}};k_{j},k_{m})=
k_{j}k_{m}\int_{0}^{L}{\rm d}\eta
\int\!\!\!\int{\rm d}^{2}\mathbf{K}
\int_{-\infty}^{\infty}{\rm d}\mu
F_{n}(\mathbf{K},|\mu|)\times
 \nonumber \\&&
\exp\left\{i\mathbf{K}\cdotp
\left[\gamma_{j}(\eta)\mathbf{r_{j}}-
\gamma_{m}^{*}(\eta)\mathbf{r_{m}}\right]
-\frac{i\kappa^{2}}{2}
\left[\frac{\gamma_{j}(\eta)}{k_{j}}-
\frac{\gamma_{m}^{*}(\eta)}{k_{m}}
\right](L-\eta)
\right\}.
\label{app27}
\end{eqnarray}
The two-dimensional spectral density is related to the 
power spectral density $\Phi_{n}(\mathbf{K})$ via
\begin{eqnarray}&&
\Phi_{n}(\mathbf{K})=
\frac{1}{2\pi}
\int_{-\infty}^{\infty}{\rm d}\mu
F_{n}(\mathbf{K},|\mu|).
\label{app28}
\end{eqnarray}
We assume that the turbulence is statistically homogeneous and
isotropic so that $\Phi_{n}(\mathbf{K})=\Phi_{n}(\kappa)$. Substituting
Eq. (\ref{app28}) into Eq. (\ref{app27}) and performing integration
over the angular coordinate in $\mathbf{K}$-space we obtain
\begin{eqnarray}&&
E_{2jm}({\mathbf r_{j}},{\mathbf r_{m}};k_{j},k_{m})=
4\pi^{2}k_{j}k_{m}\int_{0}^{L}{\rm d}\eta
\int_{0}^{\infty}{\rm d}\kappa\,
\kappa \Phi_{n}(\kappa)
J_{0}\left(\kappa\left|\gamma_{j}{\mathbf r_{j}}-
\gamma_{m}^{*}{\mathbf r_{m}}\right|\right)\times
 \nonumber \\&&
\exp\left\{
-\frac{i\kappa^{2}}{2}
\left[\frac{\gamma_{j}(\eta)}{k_{j}}-
\frac{\gamma_{m}^{*}(\eta)}{k_{m}}
\right](L-\eta)
\right\}.
\label{app30}
\end{eqnarray} 
Changing variables from $\eta$ to the normalized distance 
$\xi=1-\eta/L$ we arrive at
\begin{eqnarray}&&
E_{2jm}({\mathbf r_{j}},{\mathbf r_{m}};k_{j},k_{m})=
4\pi^{2}k_{j}k_{m}\int_{0}^{1}{\rm d}\xi
\int_{0}^{\infty}{\rm d}\kappa\,
\kappa \Phi_{n}(\kappa)
J_{0}\left(\kappa\left|\gamma_{j}{\mathbf r_{j}}-
\gamma_{m}^{*}{\mathbf r_{m}}\right|\right)\times
 \nonumber \\&&
\exp\left[
-\frac{i}{2}\kappa^{2}L
\left(\frac{\gamma_{j}}{k_{j}}-
\frac{\gamma_{m}^{*}}{k_{m}}
\right)\xi
\right],
\label{app31}
\end{eqnarray} 
which is Eq. (\ref{pcb14}). Equation (\ref{pcb15}) for 
$E_{2mj}({\mathbf r_{m}},{\mathbf r_{j}};k_{m},k_{j})$
and Eq. (\ref{pcb16}) for 
$E_{3jm}({\mathbf r_{j}},{\mathbf r_{m}};k_{j},k_{m})$
are obtained in a similar manner.

\newpage

\section*{List of Figure Captions}

Fig. 1. Longitudinal component of the scintillation index 
$\sigma^{2}_{I,l}$ as a function of the initial beam separation $d$.
The solid line is the result obtained by using the
the Von K\'arm\'an spectrum. The dashed line is the result obtained by
using the Kolmogorov spectrum and the dotted line corresponds to
the result obtained by using the Kolmogorov spectrum and assuming
that $r_{j}/W_{j}\ll 1$. The square/circle stand for the 
longitudinal scintillation of a single beam with the same total power and the
same initial spot size/amplitude, respectively.

\noindent Fig. 2. Self- and cross-intensity contributions to the longitudinal
scintillation index $\sigma^{2}_{I,l,s}$ and $\sigma^{2}_{I,l,c}$,
respectively, vs beam separation $d$. The dashed line represents
$\sigma^{2}_{I,l,s}(d;L)$ and the dotted line stands for 
$\sigma^{2}_{I,l,c}(d;L)$. The solid line corresponds to 
$\sigma^{2}_{I,l}(d;L)+1/2$.

\noindent Fig. 3. Total scintillation index  $\sigma^{2}_{I}({\mathbf r},L)$ 
for the two-beam system with the optimal configuration $d_{0}=2.8$cm at 
a propagation distance $L=1$km. The figure shows a 8cm$\times$8cm 
domain centered about the $z$ axis.

\noindent Fig. 4. Fractional reduction of the total scintillation index 
$R_{a}({\mathbf r},L)$ obtained by using the optimal two-beam system 
relative to a single Gaussian beam with the same total 
intensity and initial spot size.

\noindent Fig. 5. Self-intensity contribution to the total scintillation index
$\sigma^{2}_{I,s}({\mathbf r},L)$ for the optimal two-beam system.

\noindent Fig. 6. Cross-intensity contribution to the total scintillation 
index $\sigma^{2}_{I,c}({\mathbf r},L)$ for the optimal two-beam system.

\noindent Fig. 7. Circularly averaged radial scintillation index for the
optimal two-beam system $\sigma^{2}_{rr}$ as a function of radius $r$ 
(circles). The squares correspond to the result obtained for a single 
beam with the same power and initial spot size.

\newpage

\newpage

\newpage

\newpage

\newpage

\newpage

\newpage

\end{document}